\documentclass[preprint,floatfix]{revtex4-1}
\usepackage[T1]{fontenc}
\usepackage{color}
\usepackage{float}
\usepackage{amsmath}
\usepackage{amsthm}
\usepackage{amssymb}
\usepackage{amsfonts}
\usepackage{graphicx}
\usepackage{multirow,array}
\usepackage{qcircuit}
\definecolor{red}{rgb}{1,0.,0}

\begin{document}

\title{Application of Quantum Annealing to Nurse Scheduling Problem}
\author{Kazuki Ikeda}
\email{kazuki7131@gmail.com}
\affiliation{Department of Physics, Osaka University, Toyonaka, Osaka 5600043, Japan}
\author{Yuma Nakamura}
\email{ynakamu3@utk.edu}
\affiliation{Department of Physics and Astronomy, University of Tennessee, Knoxville, TN 37916, USA.}
\author{Travis S. Humble}
\email{humblets@ornl.gov}
\affiliation{Quantum Computing Institute, Oak Ridge National Laboratory, Oak Ridge, Tennessee 37831, USA}

\begin{abstract}
Quantum annealing is a promising heuristic method to solve combinatorial optimization problems, and efforts to quantify performance on real-world problems provide insights into how this approach may be best used in practice. We investigate the empirical performance of quantum annealing to solve the Nurse Scheduling Problem (NSP) with hard constraints using the D-Wave 2000Q quantum annealing device. NSP seeks the optimal assignment for a set of nurses to  shifts under an accompanying set of constraints on schedule and personnel. After reducing NSP to a novel Ising-type Hamiltonian, we evaluate the solution quality obtained from the D-Wave 2000Q against the constraint requirements as well as the diversity of solutions. For the test problems explored here, our results indicate that quantum annealing recovers satisfying solutions for NSP and suggests the heuristic method is sufficient for practical use. Moreover, we observe that solution quality can be greatly improved through the use of reverse annealing, in which it is possible to refine returned results by using the annealing process a second time. We compare the performance NSP using both forward and reverse annealing methods and describe how these approach might be used in practice.

\end{abstract}

\flushbottom
\maketitle

\thispagestyle{empty}

\section{Introduction}
Among non-deterministic polynomial time (NP)-hard problems, timetabling or rostering problems represent a number of practically important examples. For example, in operations research, the nurse scheduling problem (NSP) arises when finding the optimal schedule for a set of available nurses over a fixed timetable of shifts. Solutions to  NSP are required to respect hard constraints, such as days off and minimum availability, as well as soft constraints, such as minimum shift assignments, for each nurse. Examples of NSP are often cast as linear or quadratic programming problems, depending on the nature of the constraints, but they may also be formulated in terms of unconstrained optimization and solved using search methods, including tabu search.
\par 
Quantum annealing (QA) is a metaheuristic method for solving combinatorial optimization problems derived from the principles of quantum mechanics \cite{PhysRevE.58.5355}. QA operates by driving the Hamiltonian dynamics of an initial quantum state to a sought-after final state that represents the minimum energy configuration of an encoded optimization problem \cite{RevModPhys.90.015002}. In the limit that the dynamics are strictly adiabatic and the Hamiltonian sufficiently complex, this coincides with the universal model of adiabatic quantum computing \cite{Farhi472,doi:10.1137/080734479}. In practice, however, the adiabatic condition is rarely obtained and the guarantee that the true solution will be recovered is lost. Instead, these quasi-adiabatic dynamics  yield the sought-after final state with some non-unit probability and it remains open as to when such behavior can provide a computational advantage \cite{RevModPhys.90.015002,2018arXiv180309954F}. 
\par
We demonstrate the use of QA to solve NSP and we evaluate its efficiency and accuracy for this problem from empirical results. Our approach uses the commercial quantum annealer available from D-Wave Systems to implement several hard constraints. The D-Wave 2000Q is a commercially available quantum annealing device based on superconducting flux qubits designed to solve quadratic unconstrained binary optimization (QUBO) \cite{Johnson2011}. It uses the principles of QA operating in the presence of a transverse-Ising Hamiltonian by encoding the problem into a sparsely connected graph expressing the hardware interactions. By reducing NSP to QUBO form and then embedding this problem into the D-Wave processor, we use QA to recover candidate solutions for different problem instances.
\par
Previous applications of QA to unconstrained optimization span a broad variety of topics \cite{venturelli2015quantum,2018arXiv181102524B,2017arXiv170801625N,2017arXiv171104889S,2016ISTSP..10.1053R,10.3389/fict.2017.00029,king2018observation,O'Malley18}, while some efforts have investigated closely related timetabling problems, such as the job-shop problem \cite{2015arXiv150608479V}. The distinguishing features of NSP  include the multiple hard constraints, which make the problem difficult to address for conventional numerical solvers \cite{cooper1995complexity}. Equally challenging is the effort to recover approximately optimal solutions, which may fail to find the true minimum with some parameterized tolerance. 
\par 
Our interest lies in casting NSP as a combinatorial optimization problem in order to validate the capabilities of QA. This may be seen as a first step toward developing quantum-enhanced solvers for other scheduling problems as well. However, the question as to whether QA satisfies the question of quantum computational advantages cannot be addressed here. In addition to limitations on the  existing hardware, in both capacity and control \cite{2018arXiv180100862P}, there is poor theoretical understanding of the computational power for QA generally. Prior analyses have suggested that NSP and other classical combinatorial optimization problems belong to the complexity class stoqAQC, for which the relationship to the uniquely quantum complexity class BQP is not yet established \cite{RevModPhys.90.015002}. These so-called stoquastic problems may be solved efficiently by strictly classical methods, but this is also unknown \cite{2018arXiv180309954F}. However, there is a clear potential for quantum advantage for non-stoquastic problems, which do not correspond to NSP or other combinatorial optimization problems \cite{2019arXiv190306139O}.
\par 
Our presentation is organized as follows. Sec.~\ref{sec:NSP} defines NSP formally in terms of the hard and soft constraints considered here; Sec.~\ref{sec:method} introduces QA specialized to the D-Wave 2000Q in terms of a transverse Ising Hamiltonian that encodes an NSP instance. Results from evaluating various problem instances on the 2000Q processor are summarized in Sec.~\ref{sec:results}, and the influence of reverse annealing to improve these results are discussed. Lastly, we conclude in Sec.~\ref{sec:conclusion} with comments for future work.
\section{Nurse Scheduling Problem}\label{sec:NSP}
Several alternative definitions for NSP exist, yet its fundamental concepts can be summarized as follows. NSP is a problem to create a rotating roster of nurses working at a hospital while respecting constraints on their availability and level of effort. In the simplest example of a two-shift system, a schedule assigns nurses to the day duty and night duty shifts of the roster. We will apply two hard constraints to this example that enforce a minimum number of nurses assigned to each shift while also ensuring a minimal period of rest between shifts for each nurse. We refer to these as shift constraints and nurse constraints, respectively. 
\par 
Shift constraints require that a sufficient number of nurses be assigned to each shift. However, the necessary number may depend on the experience of the assigned nurses, as more experienced nurses may be capable of performing more work during their shift.  Nurses usually work as part of a group with each in charge of multiple jobs. Hence it is important to allocate the minimum number of nurses needed to cover all the jobs during a shift. In addition, sometimes it would be also required to distribute jobs to nurses evenly. By contrast, nurse constraints represent the need to satisfy the appropriate working condition for each nurse. This includes time between shifts to get enough rest as well as days off and scheduled vacations. 
We will apply the following constraints to our instance of NSP:
\begin{enumerate}
    \item Upper and lower limit of the number of breaks. 
    \item The number of nurses in duty for each shift slot. 
    \item Upper and lower limit of time interval between two days of duty.
\end{enumerate}

\section{Methods}
\label{sec:method}
\subsection{Quantum Annealing with the D-Wave 2000Q}
The D-Wave 2000Q is based on quantum annealing, which is a derivative of adiabatic quantum optimization. The latter is based on the time-dependent Schrodinger equation
\begin{equation}
    i\hbar\frac{\partial\psi(t)}{\partial t} = H(t) \psi(t)
\end{equation}
where $\psi(t)$ denotes the quantum mechanical wave function of an underlying physical system and $H(t)$ is the time-dependent Hamiltonian that drives the dynamics. A generic form of this Hamiltonian is
\begin{equation}
    H(t)=A(t)H_0+B(t)H_1,
\end{equation}
with $t\in[0,T]$ and $T$ the final evolution time. The schedules $A(t), B(t)$ are monotonic and satisfy $A(0) = 1, B(0) = 0$ and $A(T) = 0, B(T) = 1$. Therefore, the quantum state $\psi(0)$ evolves under an interpolation from $H_0$ to $H_1$ in order to prepare the final state $\psi(T)$. Assuming the initial state is an eigenstate of $H_0$, then the adiabatic theorem promises that the quantum state will remain an instantaneous eigenstate of $H(t)$ provided the dynamics evolve sufficiently slow. The latter condition may be enforced by choice of the annealing time $T$ or the schedules. Consequently, we may select the final Hamiltonian $H_1$ to represent a computational problem in which the eigenstates encode a well-defined solution. More precisely, we will focus on the case in which the ground state encodes the computational solution.
\par 
Let $s^x_i,s^z_i$ be Pauli spin operators at sites. The D-Wave 2000Q implements an initial Hamiltonian expressed as 
\begin{equation}
    H_0=-\sum_{i\in V}s^x_i,
\end{equation}
where $V$ is a set of spin sites, and a final Hamiltonian that takes the Ising spin model form
\begin{equation}
\label{eq:ising}
    H_1=\sum_{i\in V}J_{ij}s^z_is^z_j+\sum_{i\in V}{h_is^z_i},
\end{equation}
where $J_{ij}$ describe the interaction between sites $i,j$ and $h_i$ are the weights of the linear terms. The resulting time-dependent Hamiltonian corresponds to the well-known transverse field Ising model which is used widely in statistical physics to describe complex spin systems. Finding the ground state of the Ising model itself is known to be NP-Hard and, therefore, this Hamiltonian is capable of expressing a broad variety of combinatorial optimization problems including, as we show below, NSP. By contrast, the ground state of $H_0$ is easy to deduce.
\par
Notwithstanding the premise of adiabatic quantum optimization, the technical challenges of reliably preparing a quantum physical system in a pure, zero-temperature quantum state prevent these ideals from being realized in practice. Rather, the behavior of the 2000Q is better approximated by a mixed quantum state evolving as an open system. In addition to the limits on controllability, there is more general concern that knowing the optimal annealing schedule and duration require a priori information about the solution itself. Therefore, quantum annealing is most often treated as a heuristic that may be applied with good accuracy in certain situations. 
\par
In our examples below, we interface with the quantum annealer by providing a logical representation of the Ising spin model $H_1$ to be solved. Additional steps address the transformation of the logical input into a physical representation which can be embedded into the hardware \cite{humble2014integrated}. This step, known as minor embedding, depends strongly on the connectivity of the vertex set $V$ and the sparsity of the chimera layout \cite{klymko2014adiabatic,Hamilton2017}. Consequently, additional auxillary spin variables may be introduced during embedding to ensure logical connections are satisfied \cite{10.1371/journal.pone.0207827}. In addition, the ability to express the logical parameters, $J_{i,j}$ and $h_i$, is limited by the dynamic range of the hardware control.
\par 
The 2000Q offers a variety of controls for modifying the annealing time and schedules that determine the quantum dynamics leading to solution. Forward annealing corresponds to the process of driving the quantum system from the initial Hamiltonian $H_0$ to the final Hamiltonian $H_1$ and then performing measurements. We use forward annealing to compute the outcome of a given problem instance and we repeat this process many times to estimate the frequency with which computed solutions are observed. Reverse annealing builds on the result of a forward anneal by first initializing the processor to a previously computed result and then reversing the Hamiltonian dynamics to anneal backward. This process reintroduces the transverse field and potentially prepares a new, intermediate quantum state. The process is then completed by annealing forward in time to the final Hamiltonian. We investigate the relative accuracy of both forward and reverse annealing to solve instances of NSP.
\subsection{Ising Model Formulation of NSP}
A common approach to casting combinatorial optimization as an Ising model is to first express the problem as unconstrained optimization and, more specifically, as quadratic unconstrained binary optimization (QUBO) \cite{10.3389/fphy.2014.00005}. The QUBO form offers a direct mapping into the Ising model using a simple change of variable from binary to bipolar representation. We take this approach to formulate NSP as a QUBO problem with respect to minimization and then transform to the equivalent Ising model.
\par 
Consider a set of $N$ nurses labeled as $n=1,\ldots,N$ and a schedule consisting of $D$ working days  labelled as $d=1,\ldots,D$. Using the binary variable $q_{n,d}\in\{0,1\}$, let $q_{n,d} = 1$ specify the assignment of nurse $n$ to day $d$. We then consider specific instances of the shift and nurse constraints discussed above. For the hard shift constraint, we require that the schedule must ensure at least 1 nurse is assigned each working each day. For the hard nurse constraint, the schedule must ensure no nurse works two or more consecutive days, while the soft nurse constraint requires that all nurses should have approximately even work schedules.
\par 
We construct objective functions that correspond to each shift and nurse constraint and then use the sum of these terms to express the QUBO form. We introduce composite indices $i(n,d)$ and $j(n,d)$ as functions of the nurse $n$ and the day $d$. We construct the hard shift constraint by introducing a symmetric, real-valued matrix $J$ such that 
$J_{i(n,d),j(n,d+1)}=a$ and zero otherwise. The positive correlation constant $a$ enforces the shift constraint by penalizing a schedule for nurse $n$ to work two consecutive days. The resulting objective function is quadratic, i.e., $J_{i,j} q_{i}q_{j}$, and takes its minimal when the hard shift constraint is satisfied. Note that the shift constraint can be modified by changing the entries of the matrix $J$.
\par 
We express the hard nurse constraint in terms of the required workforce $W(d)$ needed on each day $d$ and the level of effort $E(n)$ available from each nurse $n$. We seek an equality solution for this constraint by introducing a quadratic function that penalizes schedules with too many or too few nurses assigned. We take a similar approach for the soft nurse constraint by introducing a quadratic penalty for failing to account for nurse preferences in the work schedule. We use $F(n)$ to specify the number of work days that each nurse wishes to be scheduled and $G(n,d)$ to define a the preference for nurse $n$ to work on day $d$.
\par 
As a simplified example of the soft nurse constraint, we decompose the preference function into the product $G(n,d) = h_1(n) h_2(d)$, in such a way that 
\begin{equation}
h_1(n)=
\begin{cases}
3h_1&\text{busy}\\
2h_1&\text{moderate}\\
h_1&\text{idle},
\end{cases}
\end{equation}
where $h_1$ is a positive value. In addition, they can also have options whether they may work on weekend/night or not by tuning $h_2(d)$:
\begin{equation}
h_2(d)=
\begin{cases}
2h_2&\text{weekend or night}\\
h_2&\text{weekday},
\end{cases}
\end{equation}   
where $h_2$ is a positive value. We simply require the minimum duty days $F(n)$ for all nurses $n$ are equal to or greater than $[D/N]$, where $[x]~(x\in\mathbb{R})$ means the integer part of $x$.
\par 
Composing these individual terms into a single objective function yields the QUBO form
\begin{align}\label{H}
\begin{aligned}
H_1(q)=&\sum_{n,n'}^{N}{\sum_{d,d'}^{D}{J_{i(n,d),j(n',d')}q_{i(n,d)}q_{j(n',d')}}}\\
&+\lambda\sum_d^D\left(\sum_n^N E(n)q_{i(n,d)}-W(d)\right)^2+\gamma\sum_n^N\left(\sum_d^D h_1(d)h_2(n)q_{i(n,d)}-F(n)\right)^2
\end{aligned}
\end{align}
where the positive real-valued numbers $\lambda$ and $\gamma$  tune the relative significance of each term. The objective function has its minimum when all the constraints are satisfied and takes on a positive value otherwise. We will assume that the functions $E(n),F(n)$ and $W(d)$ are integer-valued functions of $n$ or $d$ but this is not required. We will require the minimum duty days $F(n)$ for all nurses $n$ are equal to or greater than $[D/N]$, where $[x]~(x\in\mathbb{R})$ means the integer part of $x$. 
\par 
We next transform the QUBO expression in Eq.~(\ref{H}) into an equivalent Ising spin model. This requires changing from the binary variables $q_{i}$ to the bipolar spin variable $s_{i} = 2q_{i} - 1$. The resulting quadratic terms are then collected to match the form of the Ising spin model in Eq.~(\ref{eq:ising}). Notably, the connectivity between spin sites is determined by the relatively sparse shift constraint matrix $J$ as well as the nurse constraints set by $E(n)$, $W(d)$, $F(n)$, and $G(n,d)$. 
\section{Results}\label{sec:results}
Using the Ising model for NSP, we solve this combinatorial optimization problem with the D-Wave 2000Q. Our implementations use the randomized embedding algorithm based on work by Cai, Macready and Roy and implemented in the D-Wave software toolchain \cite{cai2014practical}. We first discuss results obtained using forward annealing and then with reverse annealing. For our studies, we fix the annealing time for a single sample to 200 $\mu s$ and we collect 1000 samples per problem instance to estimate the solution frequency. We present results for $N=3$ and $4$ nurses and $D=5-14$ days. Throughout our study, we fix the parameters $\gamma = 0.3$ and $\lambda = 1.3$, and for the soft nurse constraints, we use the simplest example with  $h_1(n)=1$, $h_2(d)=1$, $E(n)=1$, $W(d)=1$ and the penalty $a=7/2$. 
\par
\begin{figure}
\centering
\begin{minipage}{8cm}
    \includegraphics[width=8cm]{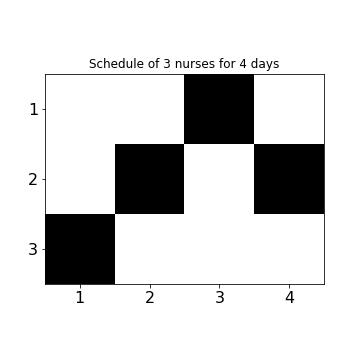}
    (a)
\end{minipage}
\centering
\begin{minipage}{8cm}
    \includegraphics[width=8cm]{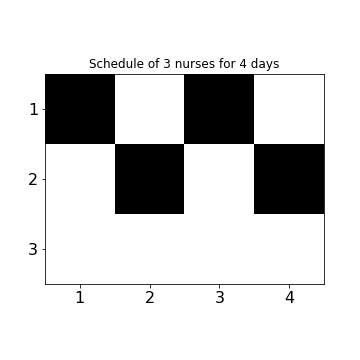}
    (b)
\end{minipage}
\caption{Graphical examples of the computed shift schedules for the case of $N = 3$ nurses for $D = 4$ days. Black squares denote a scheduled nurse who is otherwise off duty. The horizontal axes labels the days and the vertical axes labels the nurses. (a) Represents a schedule that satisfies all the constraints, while (b) represents a schedule that fails to satisfy the soft nurse constraint of every nurse being assigned some days of work.}
\label{fig:ex1}
\end{figure}

 
A graphical representation of the computed schedules are shown in Fig.~\ref{fig:ex1} of two solutions obtained from forward annealing for an instance with $N = 3$ and $D = 4$. The schedule in Fig.~\ref{fig:ex1}(a) satisfies all the constraints of this NSP instance, whereas the result in Fig.~\ref{fig:ex1}(b) does not satisfy the nurse constraint that every nurse has an assignment. Consequently, solution (a) minimizes the energy of the Ising spin model, i.e., it represents a ground state, while the result in (b) is necessarily higher due to the penalty term. Similar results are obtained for all the instances tested here, and in the following, we present statistical estimates for the frequency with which the computed schedules satisfy all the constraints. Of course, the ground state may be degenerate and multiple satisfying schedules are possible.
\subsection{Forward Annealing}
\begin{figure}[H]
    \centering
    \includegraphics[width=9cm]{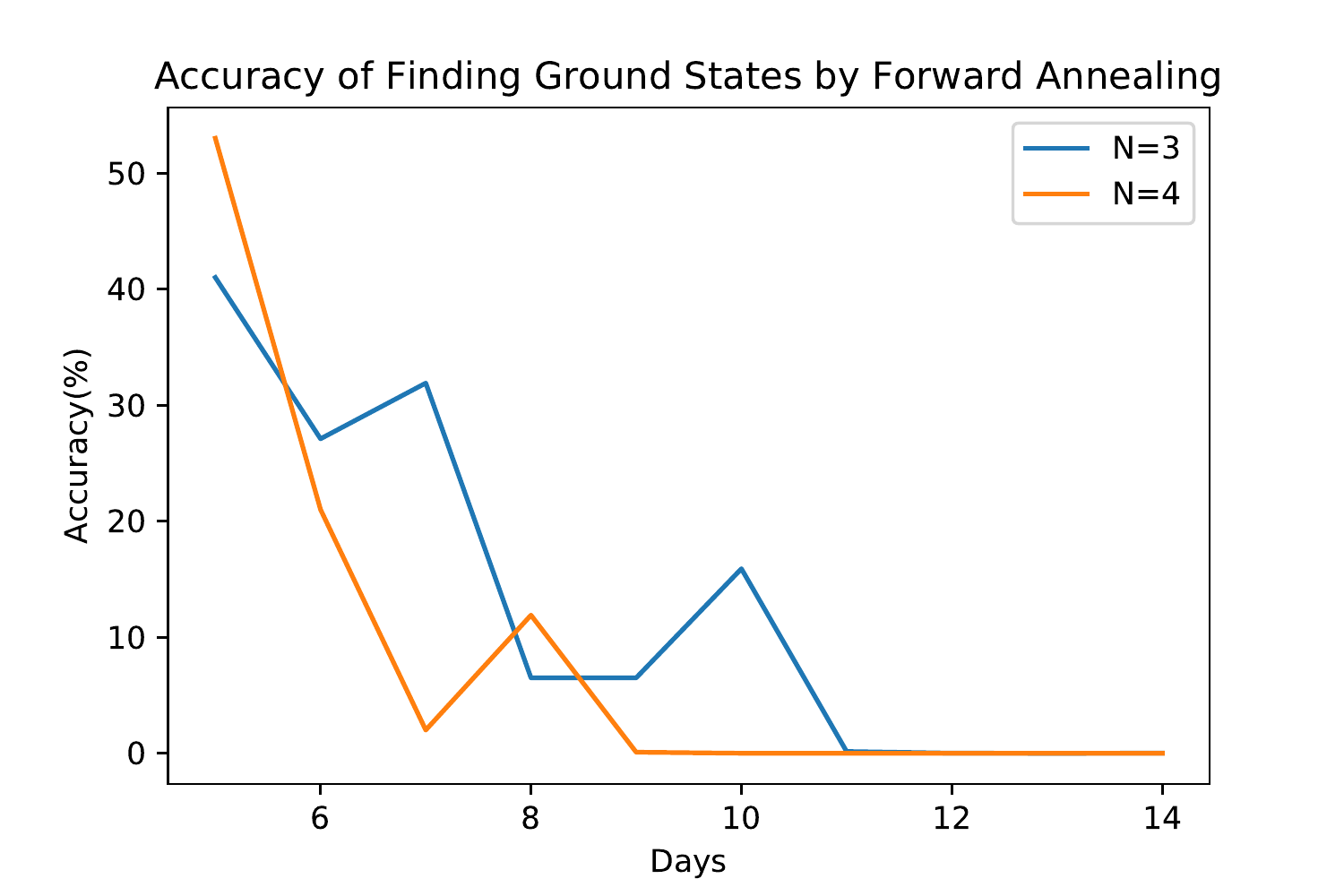}
    \caption{The frequency with which computed solutions satisfy the NSP constraints for (blue) $N=3$ and (red) $N=4$ with respect to the number of days $D$ in the schedule.}
    \label{fig:acc}
\end{figure}

We first consider the quality of solutions obtained by forward annealing. Figure \ref{fig:acc} presents the frequency with which solutions satisfy all constraints for the case of $N = 3$ and $4$ over a schedule of $D$ days. While there is significant probability to recover completely satisfying solutions for smaller schedules, we observe that the probability to satisfy all constraints falls below our sampling level for larger schedules. In particular, we did not observe completely satisfying solutions for $N=3,D=13$ and $N=4,D=10,11,12,13,14$, whereas some ground states were obtained for $N=3,D=11,12,14$ and for $N=4,D=9$. 
\par
\begin{figure}[H]
    \centering
    \includegraphics[width=9cm]{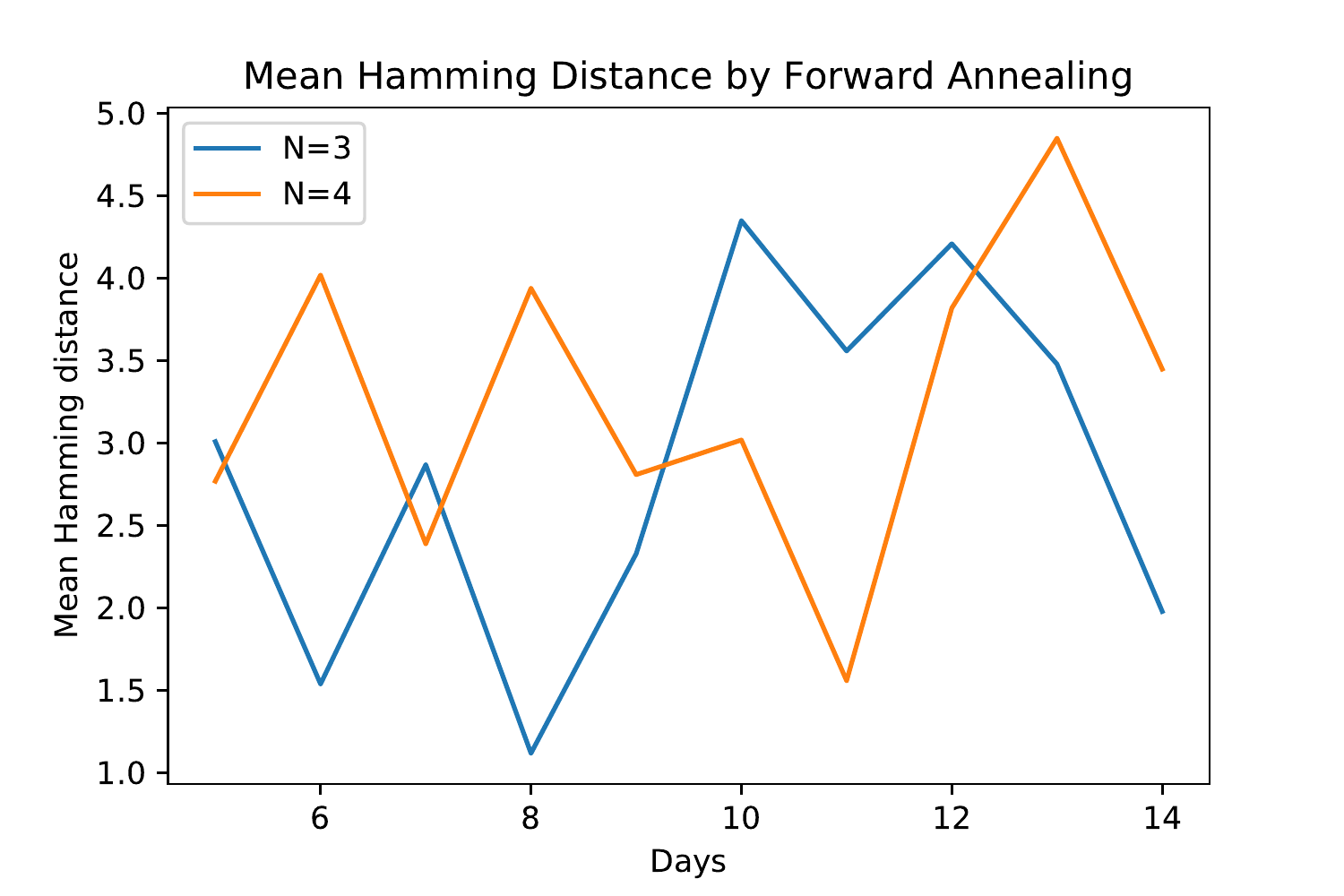}
    \caption{The mean Hamming distance of forward annealing solutions with respect to number of schedule days for (blue) $N = 3$ nurses and (red) $N = 4$ nurses.}
    \label{fig:F_meanHamm}
\end{figure}

\begin{figure}[H]
    \centering
    \includegraphics[width=9cm]{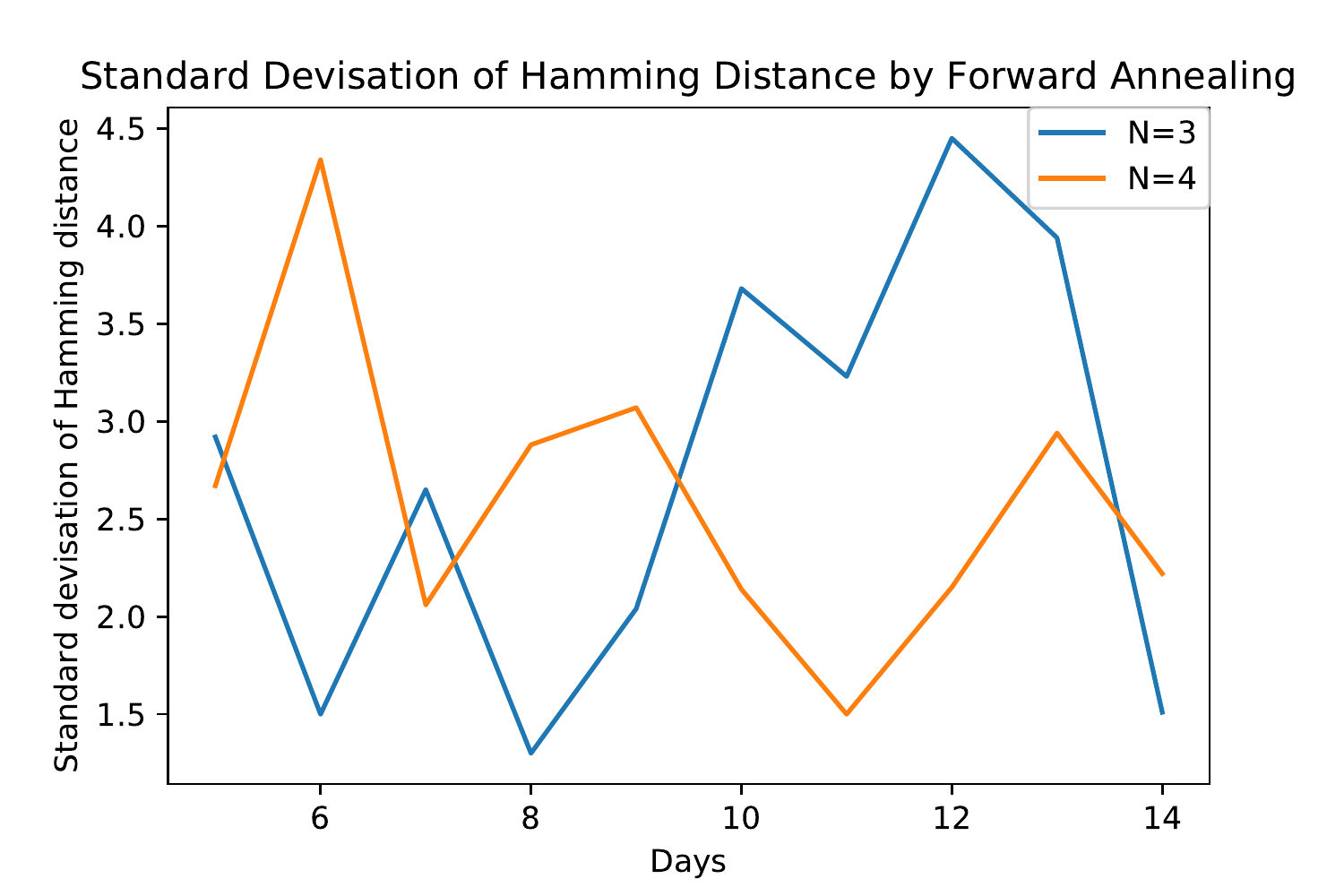}
    \caption{The standard deviation in the Hamming distance of forward annealing solutions with respect to number of schedule days for (blue) $N = 3$ nurses and (red) $N = 4$ nurses.}
    \label{fig:F_stdHamm}
\end{figure}

 
We next evaluate the deviation of computed solutions from the completely satisfying solution using the Hamming distance. For these calculations, we express the schedules as ordered binary vectors of size $ND$ and we sum the elements of their inner product to calculate the number of position in which they agree. Subtracting this number from $ND$ yields the Hamming distance, which measure the number of position in which the vectors disagree. In particular, a value of zero indicates exact agreement between the computed schedule and the observed ground state solution. Figure \ref{fig:F_meanHamm} plots the mean Hamming distance observed from all computed solutions with respect to the length of the schedule, while Fig.~\ref{fig:F_stdHamm} gives the corresponding standard deviation of the Hamming distance. It is apparent from these statistics that the average Hamming distance is well below the maximum value if $ND$ but significantly greater than 0 for all forward annealing solutions. We will present a comparison with reverse annealing solutions in the next section.
\subsection{Reverse Annealing}
We study the performance of reverse annealing when using the schedules computed by forward annealing. Reverse annealing is a heuristic methods to improve the frequency with which a computed solution satisfies the problem constraints. The procedure basically consists of the following three steps: (1) evolving the Hamiltonian $H(T)$ to an intermediate value $H(sT)$ where $s\in[0,1]$, (2) pausing the evolution for a hold period $h_t$, and then (3) evolving from $H(sT)$ to $H(T)$. The first step begins from a candidate solution and prepares an intermediate computational state of the quantum annealer, while the pause period enables the intermediate computational states to non-adiabatically mix with nearby instantaneous eigenstates due to the applied transverse field., The final step yields to potentially new computed solution state.
\par
\begin{figure}[H]
    \centering
    \includegraphics[width=8cm]{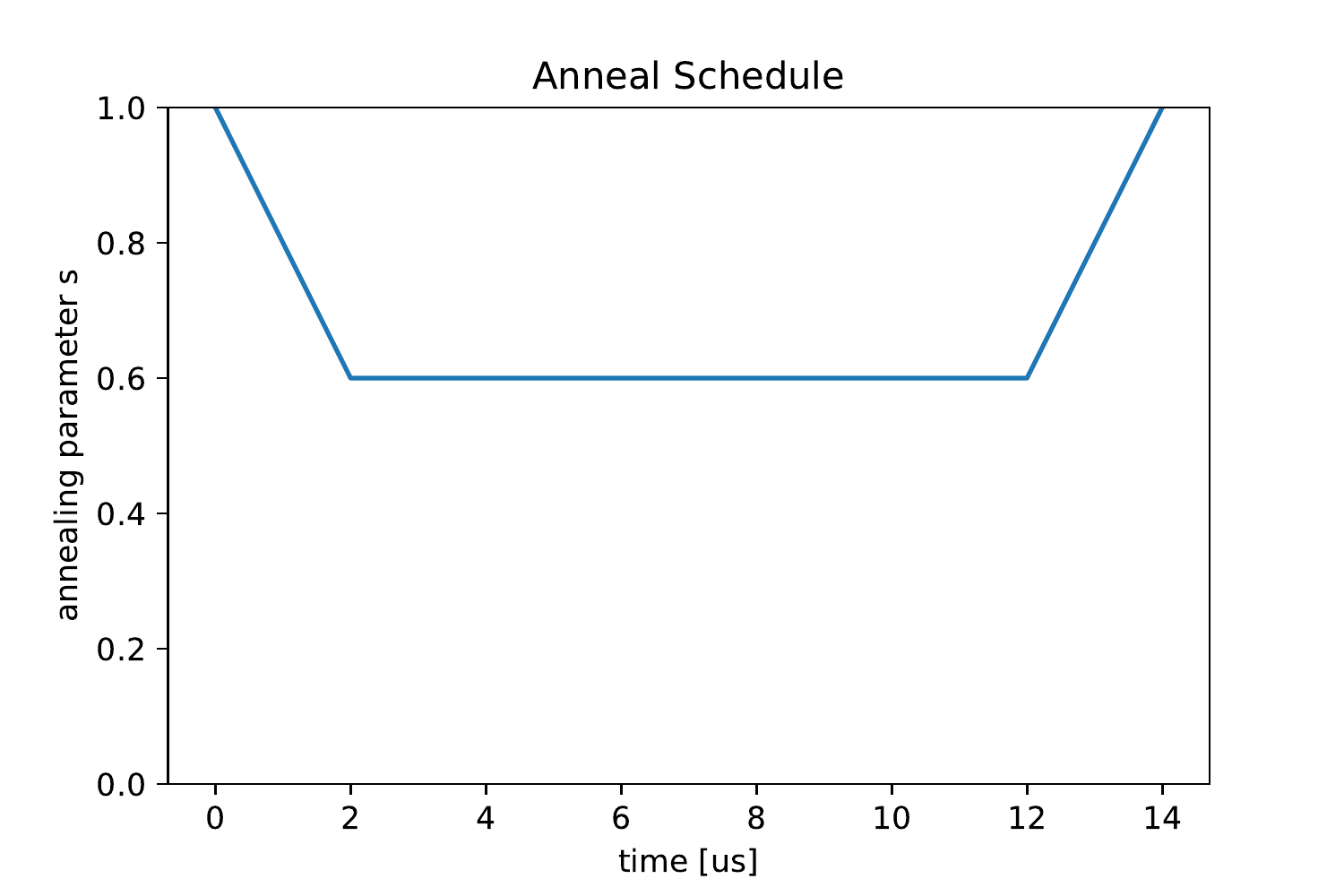}
    \caption{An example of the reverse annealing schedule used in our study plotted with respect to the evolution of the dimensionless parameter $s$.}
    \label{fig:AS}
\end{figure}

 
An example of the reverse annealing schedule used in our study is shown in Fig.~\ref{fig:AS}. The schedule reverses to $s = 0.6$ at time $2 \mu s$, holds for $h_t=10\mu s$ and then anneals forward at $12 \mu s$ to end at $14 \mu s$. Though there are various choices of the target $s$ and hold time $h_t$, we have found the heuristic to work for our problem instances when $s \in [0.6,0.8]$ and $h_t\in\{10, 50, 100 \mu s\}$. 
\par
\begin{figure}[H]
    \centering
    \includegraphics[width=8cm]{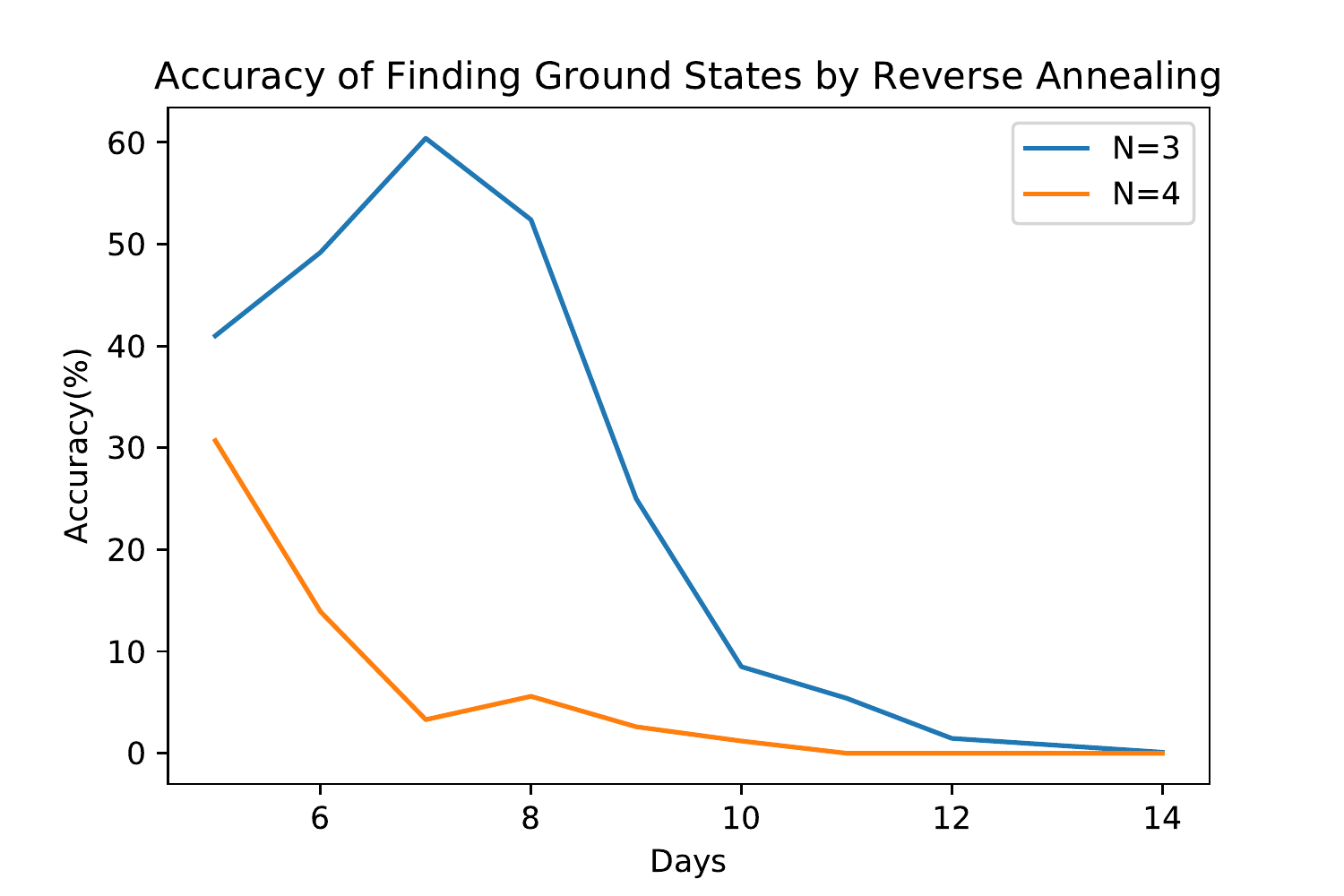}
    \caption{Accuracy of finding the ground states of the NSP Hamiltonian by reverse annealing.}
    \label{fig:R_accuracy}
\end{figure}

 
We apply reverse annealing to two different sets of forward annealing solutions. The first use of reverse annealing processes randomly chosen samples from the distribution of forward annealing solution reported in the previous section. Using the schedule shown in Fig.~\ref{fig:AS}, we show in Fig.~\ref{fig:R_accuracy} that the accuracy of finding a completely satisfying solution of the NSP instance decreases and, in some cases, vanishes. The probabilities for $N=4,D=11,12,13,14$ are exactly zero. While it is noted that solutions for $N=3,D=13$ and $N=4,D=10$ were successfully improved by reverse annealing, the average success rate decreased. By this methodology, the accuracy of finding ground states is highly related to the frequency distribution of forward annealing, since initial inputs of reverse annealing are randomly chosen. Comparing Fig.~\ref{fig:R_accuracy} with the accuracy of forward annealing given in Fig.~\ref{fig:acc}, we see that the accuracy is not always improved by reverse annealing. This is understood as follows. When accuracy is improved by reverse annealing, this is because the relatively many forward annealing solutions which are close to ground states are obtained and those solutions are frequently picked up as initial inputs of reverse annealing.  
\par
\begin{figure}[H]
    \centering
    \includegraphics[width=8cm]{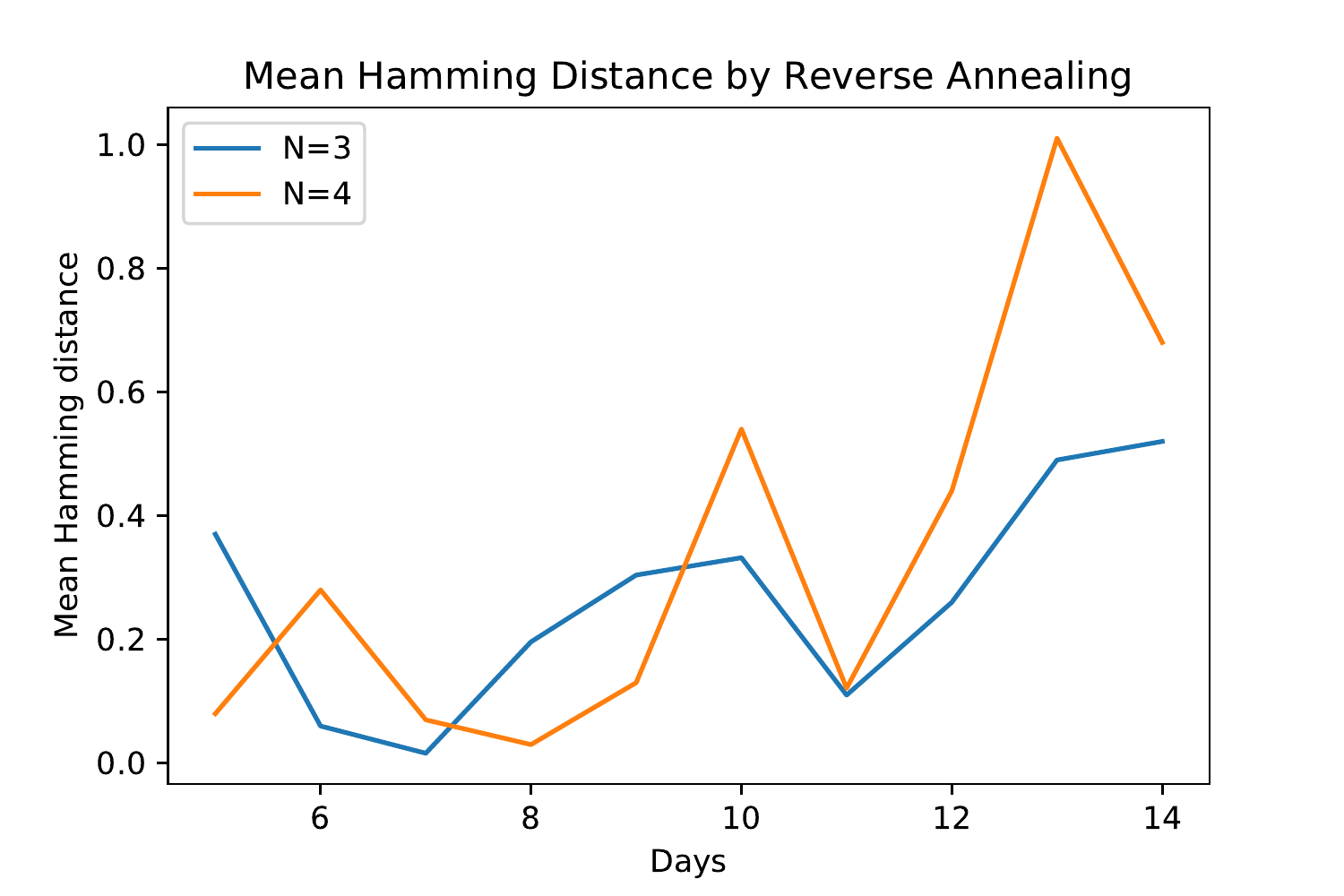}
    \caption{Mean Hamming distance of reverse annealing solutions.}
    \label{fig:R_meanHamm}
\end{figure}
\begin{figure}[H]
    \centering
    \includegraphics[width=8cm]{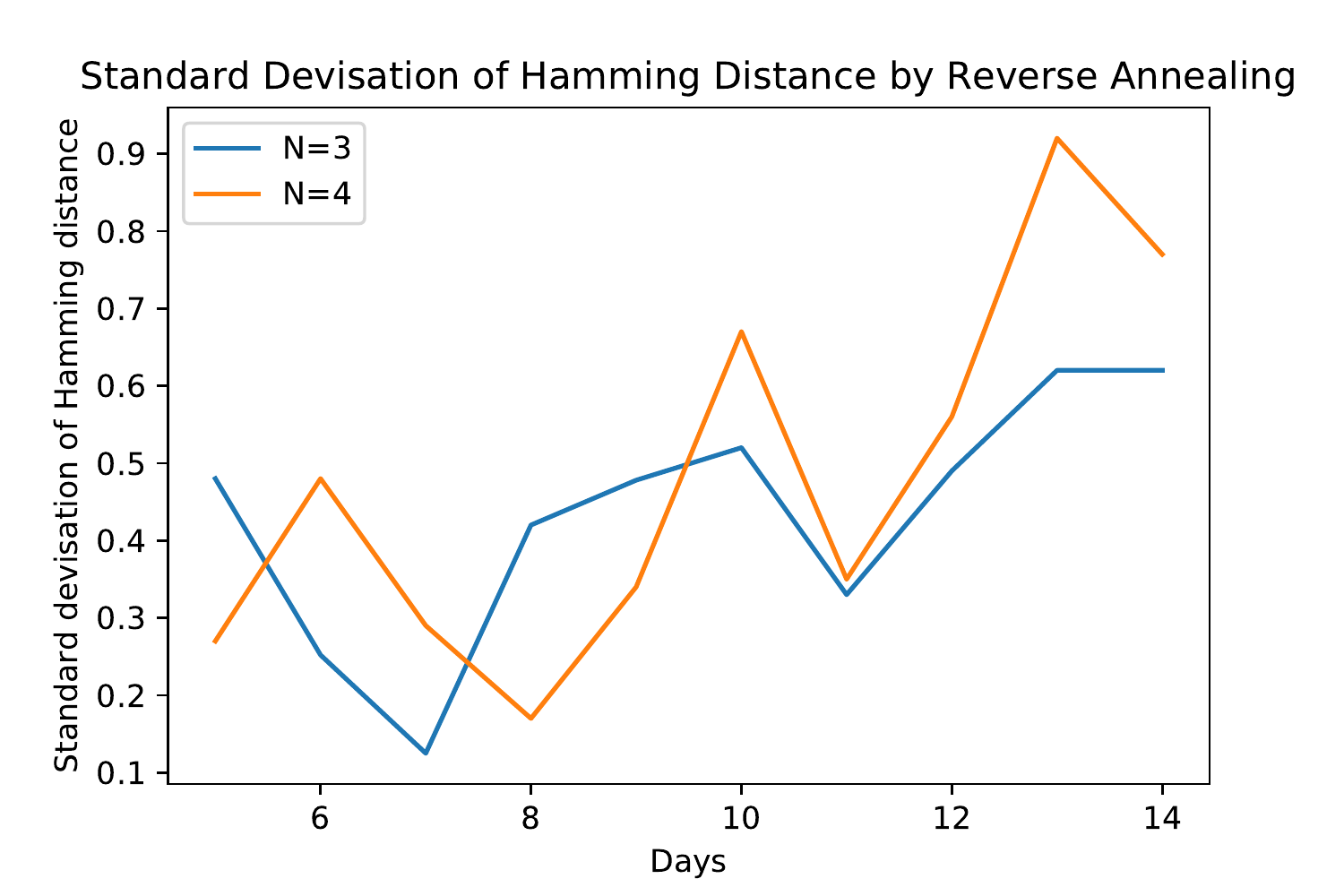}
    \caption{Standard deviation of Hamming distance of reverse annealing solutions.}
    \label{fig:R_stdHamm}
\end{figure}

We examine the corresponding Hamming distance of these solutions computed with reverse annealing in Fig.~\ref{fig:R_meanHamm}, which shows the mean Hamming distance, and Fig.~\ref{fig:R_stdHamm}, which shows the standard deviation of the Hamming distance. Comparing Fig.\ref{fig:R_meanHamm} with Fig.\ref{fig:F_meanHamm} (or Fig.\ref{fig:R_stdHamm} with Fig.\ref{fig:F_stdHamm}), we notice that overall reverse annealing decrease the mean and the standard deviation of the Hamming distance becomes close to 0, which implies that reverse annealing successfully refined solutions and grouped them together.
\par 
We now discuss application of reverse annealing to the low-lying energy states of the forward annealing solution distribution. These states represent the best solutions obtained from forward annealing and, therefore, may be the most likely to be effected by the process. This approach can be useful to estimate the essential effect of reverse annealing, since it is natural to ask whether reverse annealing works well when an initial state already close to the true ground state. 
\par
\begin{figure}[H]
    \centering
    \includegraphics[width=8cm]{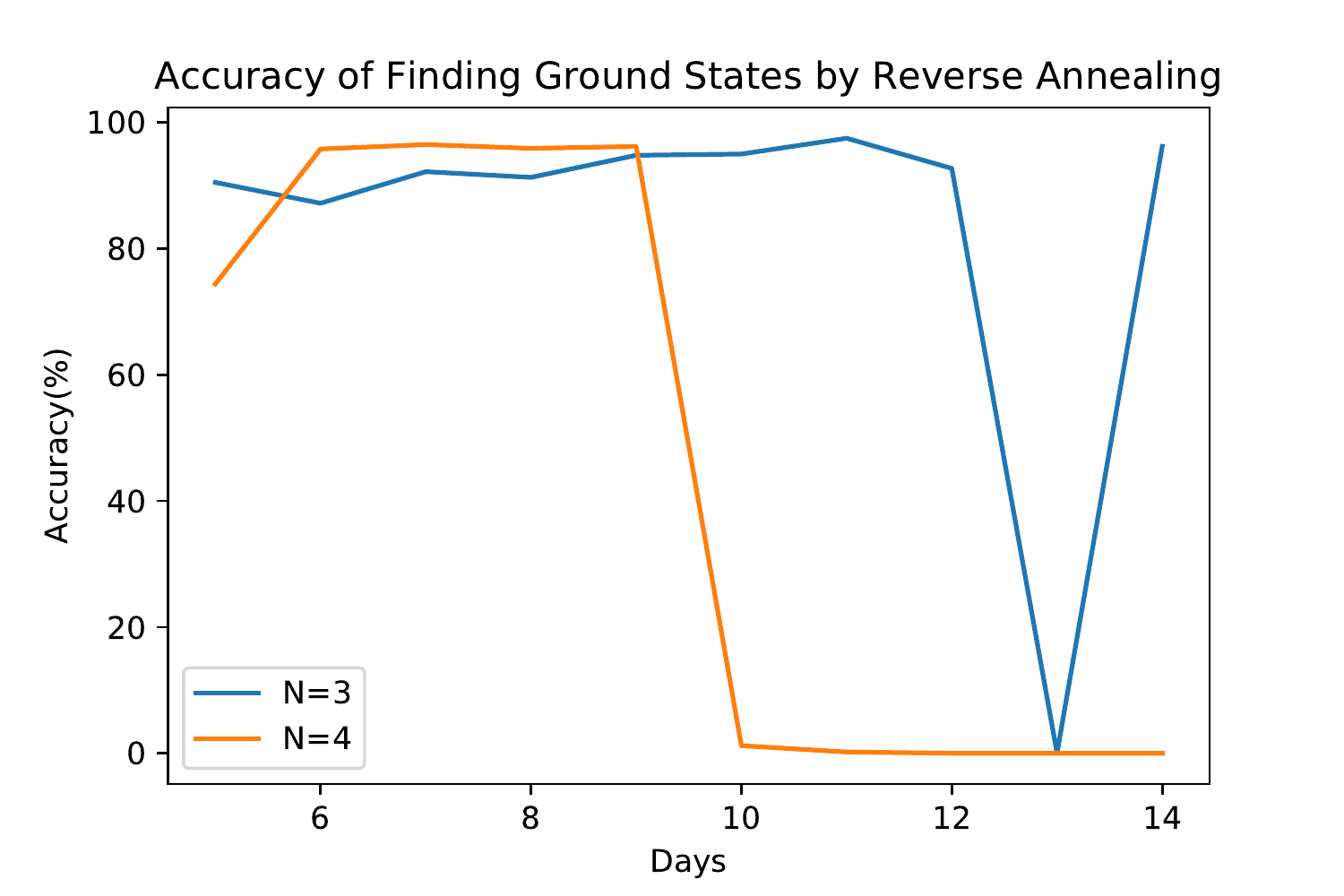}
    \caption{Accuracy of finding the ground states of the NSP Hamiltonian by reverse annealing.}
    \label{fig:R_accuracy2}
\end{figure}

 
We used the lowest energy solution from each forward annealing distribution as the initial state of reverse annealing. Figure~\ref{fig:R_accuracy2} presents the accuracy of finding a solution that completely satisfies the constraints of the NSP instance from reverse annealing with these initial states. We note that even if a satisfying solution is used as the initial input, it is not guaranteed that computed solution will be the same state. There are many local minima and reverse annealing may get trapped in such local minima depending on the schedule parameters. We find that the average accuracy at $N=3,D=13$ is exactly zero whereas at $N=3,D=14$ the accuracy is close to 100\%. These observations are consistent with the result of forward annealing data. 
\par
\begin{figure}[H]
    \centering
    \includegraphics[width=8cm]{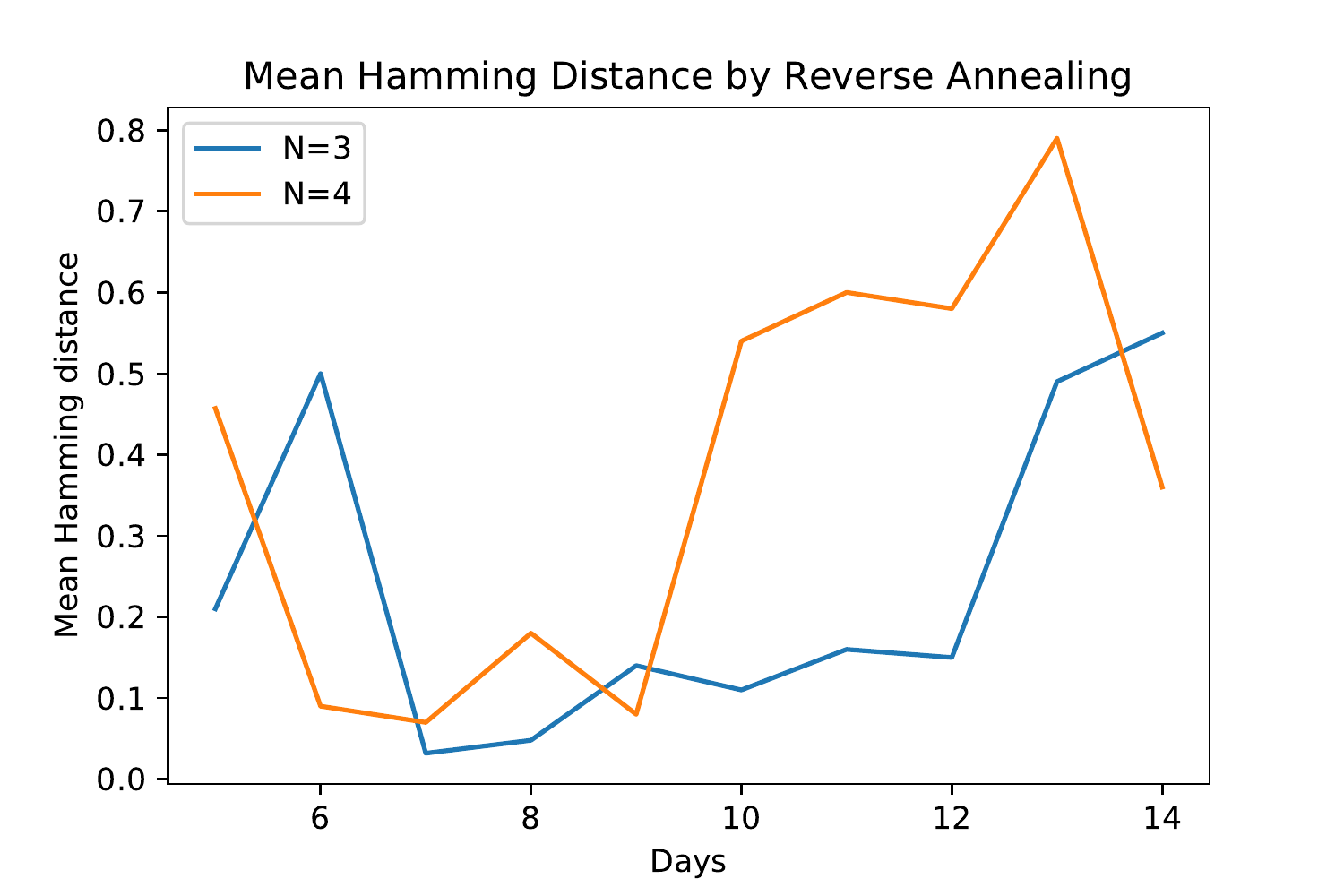}
    \caption{Mean Hamming distance of reverse annealing solutions.}
    \label{fig:R_meanHamm2}
\end{figure}

\begin{figure}[H]
    \centering
    \includegraphics[width=8cm]{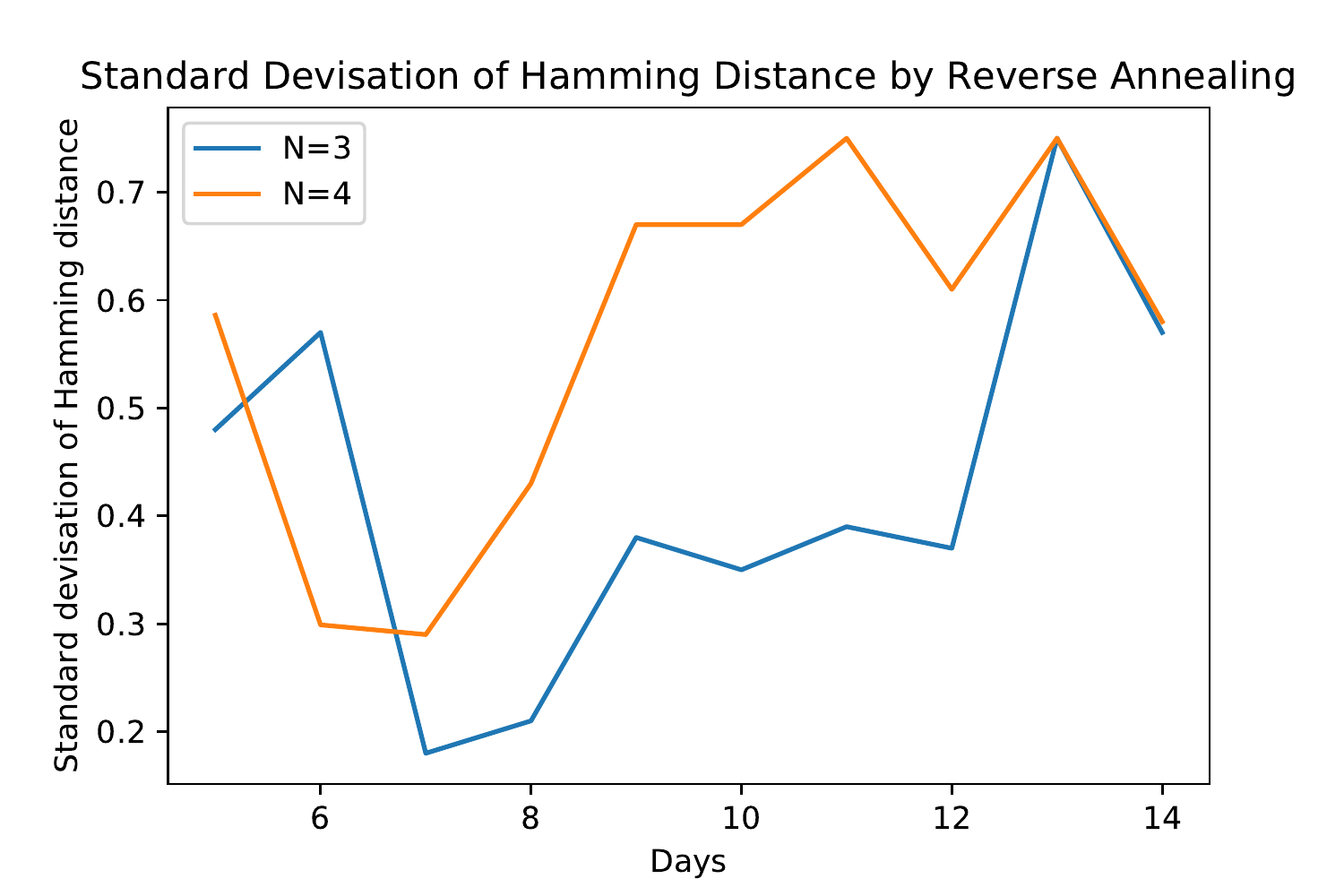}
    \caption{Standard deviation of Hamming distance of reverse annealing solutions.}
    \label{fig:R_stdHamm2}
\end{figure}

 
In Fig.~\ref{fig:R_accuracy2}, the accuracy at $N=4,D=10$ is not zero as some ground states were successfully obtained by reverse annealing even though the initial input was not a ground state. This shows that it is possible to improve solution quality by reverse annealing for the case of using the low-lying energy states as input. Figures~\ref{fig:R_meanHamm2} and \ref{fig:R_stdHamm2} show the mean Hamming distance and  standard deviation for these results. Both statistics are greatly reduced compared to the forward annealing results as expected from the improved accuracy.
\if{
A lower value of s\_target corresponds to a larger transverse field, which leads to a broader search of the solution space. A greater value of hold\_time (microseconds) implies that the reverse anneal is given more time to explore, which leads to a broader and deeper search of the solution space.
\begin{figure*}
    \centering
    \includegraphics[width=12cm]{3_10_s.pdf}
    \centering
    \includegraphics[width=12cm]{3_11_s.pdf}
    \centering
    \includegraphics[width=12cm]{3_12_s.pdf}
    \centering
    \includegraphics[width=12cm]{3_13_s.pdf}
    \centering
    \includegraphics[width=12cm]{3_14_s.pdf}
    \centering
    \includegraphics[width=12cm]{3_15_s.pdf}
    \caption{The figures show the mean Hamming distance, the mean energy and the standard deviation with respect to reverse annealing parameters for the problems $N=3$ and $D=10,\cdots,15$. Each data is based on 1000 samplings.}
    \label{3N}
\end{figure*}

\begin{figure*}
    \centering
    \includegraphics[width=12cm]{4_10_s.pdf}
    \centering
    \includegraphics[width=12cm]{4_11_s.pdf}
    \centering
    \includegraphics[width=12cm]{4_12_s.pdf}
    \centering
    \includegraphics[width=12cm]{4_13_s.pdf}
    \centering
    \includegraphics[width=12cm]{4_14_s.pdf}
    \centering
    \includegraphics[width=12cm]{4_15_s.pdf}
    \caption{The figures show the mean Hamming distance, the mean energy and the standard deviation with respect to reverse annealing parameters for the problems $N=4$ and $D=10,\cdots,15$. Each data is based on 1000 samplings.}
    \label{4N}
\end{figure*}
 }\fi

\section{Discussion and Conclusion}\label{sec:conclusion}
We have presented a formulation of the nurse scheduling problem (NSP) as an Ising spin Hamiltonian that can be solved using quantum annealing with the commercially available quantum processor. Our approach casts the constraint satisfaction problem into QUBO form and then translates directly to the Ising model. We estimate the frequency for satisfying all constraints using a variety of quantum annealing techniques performed on the D-Wave 2000Q quantum annealer. Our results demonstrate that for a fixed sample size and fixed annealing duration $T$ the probability of successfully matching all constraints decays with increasing size of the schedule $D$ and increasing size of the roster $N$. Reverse annealing methods do improve the probability of success but not uniformly.
\par 
This is an encouraging sign that the quantum annealer is suited to solve a complementary set of practical problems but we do not address the question of performance relative to conventional solver method. Current realization of quatnum annealers are insufficient in capacity to explore realistic sizes of $N$ and $D$. However, it may prove possible to decompose larger problems into a set of smaller optimization problems that could then be solved \cite{okada2019improving}. For example, we have tested an open-source software package called \textit{qbsolv}, which is based on a variant of tabu search. We found that qbsolv was capable of find satisfying solutions to NSP instances of size $(N, D)=(4, 160)$ using the same parameters used in this work.
\par 
Though quantum computation is in an early stage, the approach we have presented may be extended to more general scheduling problems, and anticipate several future directions to improve this methodology. While we have confirmed that reverse annealing can be useful to obtain ground state solutions, it is unclear the necessary parameters to ensure this improvement and further empirical investigations are warranted. In general, reverse annealing depends on several schedule variables and it is natural to try to investigate more on their dependence on those parameters. While our study has been limited to the schedule shown in Fig.\ref{fig:AS}, different schedule parameters may influence the results. An example of such a study is given in \cite{2019arXiv190204709K}. Our work is a first step toward solving scheduling problems with a quantum annealer. There would be various ways to extend and generalize our method on the NSP to different scheduling problems.

\section*{Acknowledgements}
The authors acknowledge Akira Endo for useful discussion at the initial stage of this work. We also thank 
Joel Gottlieb for carefully reading our manuscript and giving us useful comments. KI was partly supported by Scholarship of Overseas Study 2018 (Osaka University) and by Grant-in-Aid for Scientific Research on Innovative Areas, the Ministry of Education, Culture, Sports, Science and Technology, No. 19J11073. TSH acknowledges support from the Department of Energy Office of Science Early Career Research Program. Access to the D-Wave 2000Q is provided by  Oak Ridge National Laboratory. This manuscript has been authored by UT-Battelle, LLC under Contract No. DE-AC05-00OR22725 with the U.S. Department of Energy. The United States Government retains and the publisher, by accepting the article for publication, acknowledges that the United States Government retains a non-exclusive, paid-up, irrevocable, world-wide license to publish or reproduce the published form of this manuscript, or allow others to do so, for United States Government purposes. The Department of Energy will provide public access to these results of federally sponsored research in accordance with the DOE Public Access Plan. (http://energy.gov/downloads/doe-public-access-plan).

\section*{Author Contributions}
KI conducted the research, prepared the manuscript, wrote codes, corrected and analyzed all data. YN wrote codes and prepared appendix. TSH supervised the research and prepared the manuscript. 

\section*{Competing Interests}
 The authors declare no competing interests.

\section*{Appendix: Three-shift system and additional constraints}
Our example of a two-shift NSP can be extended to address a three-shift system in which each day is divided into shifts labeled as daytime, early nighttime and late nighttime. Both the two-shift and three-shift systems are popular in medical clinic settings. For the three-shift scheduled, the Hamiltonian is modified to include the new parameter $h_2(d)=\alpha (d)h'_2(d)$ as follows:
\begin{equation}
\alpha(d)=
\begin{cases}
2 &\text{weekend} \\
1 &\text{weekday},
\end{cases}
\end{equation}  
and
\begin{equation}
h'_2(d)=
\begin{cases}
2h_2 &\text{late nighttime} \\
1.5h_2 &\text{early nighttime}\\
h_2 &\text{daytime}
\end{cases}
\end{equation}
The number of duty on late nighttime shift should be finely tuned to sustain a healthy work environment of nurses. In addition, the requests of day-off for individual nurse can be incorporated to the Hamiltonian as an external magnetic field term,
\begin{align}
H=&\sum_{n,n'}\sum_{d,d'}J_{[n,d][n',d']}q_{n,d}q_{n',d'} 
+\lambda\sum_d^D\left(\sum_n^N E(n)q_{n,d}-W(d)\right)^2\\ \nonumber &+\gamma\sum_n^N\left(\sum_d^D  h_1(d)h'_2(n)q_{n,d}-F(n)\right)^2 + \eta \sum_{n,d}g(n,d)q_{n,d},
\end{align}
where $g(n,d)$ is defined by the priority of the day-off request for a specific nurse and working shift
\begin{equation}
g(n,d)=
\begin{cases}
2g &\text{high priority} \\
1.5g &\text{middle priority}\\
g &\text{low priority}
\end{cases}
\end{equation} 
By tuning the positive parameter $\eta$, the day-off request is highly considered within scheduling. This constraint should be treated as a soft constraint, whereas the first two terms correspond to hard constraints.
\par
\begin{figure*}[htbp!]
\centering    
\includegraphics[width=1.0\textwidth]{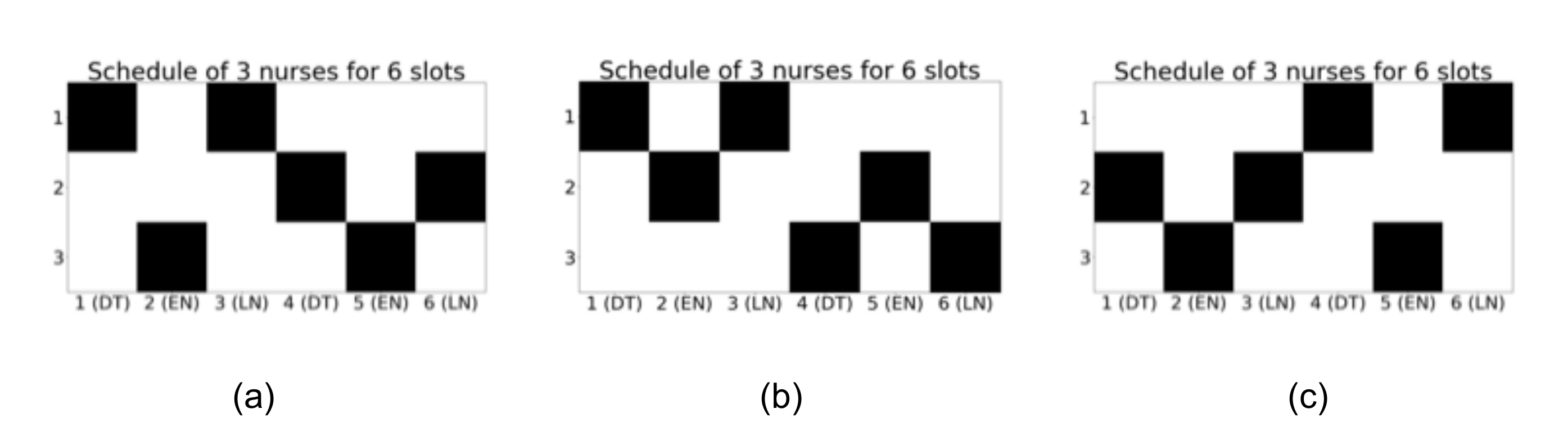}
\caption[Concept]{Solutions of the proposed three-shift NSP with additional constraints. DT, EN and LN refer to daytime, early night and late night work shift. Among (a)-(c), the day-off requests are tuned differently as $g(n,d)=0 $ for (a),$g(1,4)=1, g(2,6)=1.5$ and $g(3,5)=2$ in (b) and $g(1,6)=1, g(2,6)=1.5$ and $g(3,4)=2$ in (c)}. 
\label{fig:3}
\end{figure*}

To demonstrate that this formulation is capable of solving a three-shift NSP, we optimized the NSP with 3 nurses and 6 working slots in a three-shift system by taking the day-off request of nurse into account. In Fig. \ref{fig:3}, three different types of day-off request are incorporated for each optimization. While Fig.\ref{fig:3}(a) has no day-off request, Figs.\ref{fig:3}(b) and (c) show the job shifts according to the day-off request. In Fig.\ref{fig:3}(b), all day-off request $(n, d)$=(1,4), (2,6) and  (3,5) are satisfied. On the other hand, in Fig.\ref{fig:3}(c), not all requests are satisfied and some constraints are given priority over the day-off requests of lowest priority. As in Table \ref{tab:3}, setting the parameter $\eta$ lowest among other parameters leads day-off request as most soft constraints. The optimization is achieved by QUBO and its solutions are guaranteed to be a ground state of the Hamiltonian by the brute force search method.

 \begin{table*}[!htbp]
  \centering  
 \begin{tabular}{|l|c|c|c|c|}
    \hline
 $\lambda$& $\gamma$&$ \eta$ &W&E\\
 \hline
    1.3&0.3&0.2&1&1\\
    \hline 
 \end{tabular}
 \caption{Benchmark parameters of additional calculation}
 \label{tab:3}
\end{table*}

\bibliographystyle{apsrev4-1.bst}
\bibliography{references}

\end{document}